\documentclass[prl,aps,showpacs,twocolumn,floats,amsmath,amssymb,nofootinbib]{revtex4}

\usepackage{graphicx}% Include figure files
\usepackage{dcolumn}% Align table columns on decimal point
\usepackage{bm}% bold math

\def \be{\begin{equation}}
\def \ee{\end{equation}}

\def\nanoLLG{NanoTorque }
\begin{document}

\title{Current induced Spin Torque in a nanomagnet}

\author{X. Waintal}
\affiliation{Service de Physique de l'{\'E}tat Condens{\'e},
CEA Saclay, 
%Centre d'{\'e}tude de Saclay F-
91191 Gif-sur-Yvette cedex, France }

\author{O. Parcollet }
\affiliation{ Service de Physique Th{\'e}orique, CEA Saclay, 91191 Gif-sur-Yvette cedex, France  }

\date{\today}

\begin{abstract}
In a nanomagnet %(whose total spin $S_0\le 1000$), 
very small polarized currents can lead to
magnetic reversal. Treating on the same footing the transport and magnetic properties of a
nanomagnet connected to magnetic leads via tunneling barriers, we derive a closed equation for the time 
evolution of the magnetization. The interplay between Coulomb blockade phenomena and 
magnetism gives some additional structure to the current induced spin torque.
In addition to the possibility of stabilizing uniform spin precession states, we find that 
the system is highly hysteretic: up to three different magnetic states
can be simultaneously stable in one region of the parameter space  
(magnetic field and bias voltage).% space.
\end{abstract}

\maketitle
%%%%%%%%%%%%%%%%%%%%%%%%%%%%%%%%%%%%%%%%%%%%%%%%%%%%%%%%%%%%%%%%%%%%%%%
% INTRODUCTION
%%%%%%%%%%%%%%%%%%%%%%%%%%%%%%%%%%%%%%%%%%%%%%%%%%%%%%%%%%%%%%%%%%%%%%%

Nanomagnets (typically $3-4 \ nm$ thick grains of magnetic materials
with a total spin $S_0\sim 1000$) are model systems for the study of
the interplay between transport and magnetism.  Their magnetization
can be considered as a large macrospin without spatial
variations~\cite{jamet2001} (the magnetic exchange length is of the
order of 7 nm in cobalt so that degenerate magnons with non-zero
momentum can be ignored).
Moreover, their mean level spacing is of the
order of $1 meV$,  bigger than temperatures accessible experimentally so that
the discreteness of the grain spectrum can be
resolved~\cite{gueron1999}. A first set of experiments used single
electron tunneling spectroscopy (connecting the grain to electrodes
with tunnel junctions) to probe the grain's transport
properties~\cite{gueron1999,deshmukh2001}. The differential
conductance-voltage characteristic of such a device displays peaks
corresponding to excitations of the grain from which the grain's
spectrum can be measured~\cite{kleff2001,kleff2002}.  In a second set
of experiments the magnetization of the grain is measured directly
using a microSQUID technique ~\cite{jamet2001,thirion2003}. Reversal
of magnetic moments as small as 1000 $\mu_B$ have been observed providing
information on the static (switching field as a function of the
external field direction that allowed the reconstruction of the
anisotropy tensor) and dynamical (switching times) magnetic
properties of the system.

Magnetic reversal by injection of a spin-polarized current was
predicted in 1996 by Slonczewski~\cite{slonczewski1996} and observed
later in magnetic multilayers~\cite{katine2000}.  A very large
interest for these systems has emerged since, due in particular to
their large potential for practical applications (including current
driven Magnetic Random Access Memories and Radio Frequency
components) motivating works on magnetic tunnel junctions~\cite{slonczewski2005,fuchs2004,huai2004}. 
The spin dynamics in those systems is usually modelized with the Landau-Lifshitz-Gilbert 
(LLG) equation to which a spin-torque term is added~\cite{slonczewski1996}.

\begin{figure}
%\vglue +0.45cm
\includegraphics[width=6cm]{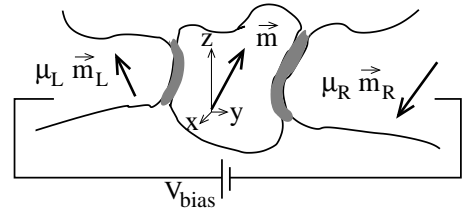}
\caption{\label{fig:system} 
Nanomagnet connected through tunneling barriers to two (magnetic or not) leads with 
chemical potential $\mu_i$ and magnetization directions $\vec m_i$ (in arbitrary directions).
The grain magnetization direction is $\vec m$ ($||\vec m|| =1$) and its easy axis $z$. The magnetic field $H$ is along $z$.}
\end{figure}

In this letter, we study the spin dynamics of a nanomagnet
connected to magnetic leads through tunneling contacts (see Fig.~\ref{fig:system}), 
in the sequential tunneling regime.
We show that the Coulomb blockade \cite{beenakker1991}
strongly affect the spin dynamics in this system. As a consequence, 
the standard LLG equation has to be replaced by %by another equation for the spin dynamics, 
Eq.(\ref{eq:rocks}) which is the main result of this letter.
In the following, we first outline the derivation of this equation,
then compare it to the standard LLG equation and study its phase diagram as
a function of the bias voltage $V_{\rm bias}$ and magnetic field $H$.
We find that highly hysteretic current-voltage
characteristics and metastable magnetic precessional states could be observed in those
systems..

Due to the interplay of Coulomb effect and spin dynamics, magnetism and transport 
have to be treated on the same footing in the derivation of the equation governing 
the spin dynamics. We first consider  
a simple model for the isolated nanomagnet, {\it i.e.} a mean-field picture of the Coulomb blockade,
a Stoner instability to get a magnet and a phenomenological modelisation of 
its intrinsic relaxation processes; we then couple the nanomagnet to the leads 
and extract the spin dynamics at lowest order in this coupling using a Master equation approach. A closely related model has been studied in~\cite{braun2004}
when the exchange interaction is below the Stoner instability.  The
Hamiltonian of the isolated grain has been previously analyzed
in~\cite{canali2000,kleff2001,waintal2003} and reads : 
\be
\label{eq:model} H_g= \sum_{\alpha\sigma} \epsilon_\alpha
d^\dagger_{\alpha \sigma } d_{\alpha \sigma } + E(N) 
- J \vec S\cdot\vec S -\frac{\kappa}{S_0} S_{z}^{2} -\hbar\gamma
H S_z 
\ee 
where $d^\dagger_{\alpha \sigma }$ creates an electron on a
one-body level $\alpha$ with energy $\epsilon_\alpha$, $N$ is the
number of electrons in the grain, $E(N) = E_C(N-N_{\rm g})^2$, with
$E_C$ the charging energy and $N_g$ an offset (controlled by a gate
voltage), $\gamma$ is the gyromagnetic ratio, $H$ the magnetic field.
The exchange energy ($\propto J$) and uniaxial anisotropy
($\propto\kappa$) depend on the total spin $\vec S = \frac{1}{2}
\sum_{\alpha,\sigma_1,\sigma_2} d^{\dagger}_{\alpha\sigma_1}
\vec\sigma_{\sigma_1\sigma_2} d_{\alpha\sigma_2}$ ($\vec\sigma$ are
Pauli matrices). Upon increasing $J$ the grain undergoes a Stoner
instability and acquires a macroscopic spin $S_0$ (equal to half of
the number of singly occupied one-body states).  Its low energy
excitations $|A\rangle$ are entirely characterized by the set of occupation number $n_\alpha= 0,1,2$ of the single energy levels
together with the spin $S_z$ along the easy
axis~\cite{canali2000,kleff2001,waintal2003}.  Intrinsic relaxation
processes (like spin-orbit or phonon-magnon scattering) are described
phenomenologically via a weak coupling $\Gamma_B$ of the form
$S^+\phi$ to a bosonic bath $\phi$ of spectral density $\rho_B(E)$.

The grain is then coupled via tunneling barriers to magnetic leads and
we note $\Gamma^{\sigma}_a$ the tunneling rate of an electron with
spin $\sigma$ ($\uparrow$ or $\downarrow$ along the $z$ axis) coming
from lead $a$ (Left or Right).  For simplicity, we restrict ourselves
to tunneling events occurring on only one one-body state $\alpha_0$,
{\it i.e.} to one step in the current-voltage Coulomb staircase ($k_B
T\ll \Delta$, where $\Delta$ is the mean level spacing).  For larger
systems where more levels come into play, spin accumulation effects
have to be taken into account~\cite{inoue2004}.  This state can be a
majority level (i.e unoccupied, $n_{\alpha_0}=0$) or a minority level
($n_{\alpha_0}=1$).  The different tunneling events (or interactions
with the bosonic bath) induce a random walk in the Hilbert space of
the isolated grain, its inner state $|A\rangle$ changing at each
event. This dynamics is described by the Master equation for the
probability $P_A(t)$ for the system to be in state $|A\rangle$ at time
$t$. The first part is a direct generalization~\cite{waintal2003} of
the standard theory of sequential tunneling~\cite{beenakker1991},
while the second part is due to the coupling to the bosonic bath :
  
\begin{widetext}
  \begin{multline}
    \label{eq:master}
%\nonumber
\partial_t P_A=
%  \frac{\partial P_A}{\partial t}=
\sum_{A',a,\sigma }%= \uparrow,\downarrow}
\Biggl \{ \Gamma_a^\sigma \left| \langle A' |
  d^{\vphantom{\dagger}}_{\alpha_0\sigma} | A \rangle \right|^2
\biggl[ n\bigl(\Delta E-\mu_a\bigr) P_{A'} - \Bigl( 1 - n\bigl(\Delta
E-\mu_a\bigr) \Bigr) P_{A} \biggr] + \Gamma_a^\sigma \left|\langle A'
  | d^{\dagger}_{\alpha_0\sigma} | A \rangle \right|^2 \biggl[
-n(-\Delta E-\mu_a) P_{A}
\\
+ \Bigl(1 - n\bigl(-\Delta E-\mu_a\bigr)\Bigr) P_{A'} \biggr] \Biggr\}
+ \Gamma_B \!\!\sum_{A',\epsilon = \pm} \!\!\epsilon \left|\langle A'
  | S_x-i\epsilon S_y | A \rangle \right|^2 \biggl[ \rho_B(\epsilon
\Delta E) n_B(\Delta E) P_{A'} + \rho_B(\epsilon\Delta E) n_B(-\Delta
E) P_A \biggr]
  \end{multline}
\end{widetext}
where $\Delta E = E_A - E_{A'}$ is the energy difference between state
$|A\rangle$ and $|A'\rangle$, $n(E)=1/(1+e^{E/kT})$
($n_B(E)=1/(e^{E/kT}-1)$) is the Fermi (Bose) function, $T$ the
temperature and $\mu_a$ the chemical potential of lead $a$.  For
Eq.(\ref{eq:master}) to be valid, the different tunneling events (and
interactions with the bosonic bath) have to be incoherent, which
requires $\Gamma,\Gamma_B \ll kT/\hbar$.  In addition,
Eq.(\ref{eq:master}) does not include the dynamics of the
magnetization in the $xy$ plane (off-diagonal terms of the density
matrix).  This is valid when internal precession is fast enough to
average out contributions transverse to the easy axis. The time needed
for $S_z$ to change significantly is of order $S_0/\Gamma$ (one
tunneling event can only change $S_z$ by half a unit) so that this
condition implies $\Gamma\ll\kappa S_0/\hbar$.  In this paper, we
restrict ourselves to that case.  The effective tunneling rates
$\Gamma_{a}^{\sigma}$ along the easy axis $z$ can be related to the
tunneling rates for electrons polarized along the magnetization of the
leads $\tilde\Gamma_{a}^{ \beta},\beta=U,D$ (Up,Down) through
$\Gamma_{a}^{\uparrow}=\tilde\Gamma_{a}^{U}\cos^2 (\theta_a/2)
+\tilde\Gamma_{a}^{D}\sin^2 (\theta_a/2)$ and $\Gamma_{a}^{
  \downarrow}=\tilde\Gamma_{a}^{D}\cos^2 (\theta_a/2)
+\tilde\Gamma_{a}^{U}\sin^2 (\theta_a/2)$ where $\theta_a$ is the
angle between the $z$ axis and the magnetization of lead $a$.  The
$\tilde\Gamma_{a}^{ \beta}$ are given by the Fermi golden rule.  A
complete derivation of Eq.(\ref{eq:master}) from the microscopic model
will be provided elsewhere~\cite{ParcolletWaintalLong}.  Since $S_0$
is macroscopic, we proceed with expanding Eq.(\ref{eq:master}) with
respect to $1/S_0$, introducing the continuous variable $m_z=S_z/S_0$.
To zeroth order in $1/S_0$ the magnetization does not change and the
charge equilibrium (between $N_0$ and $N_0+1$ electrons in the grain)
takes place.  Denoting by $P_+(m_z)$ (resp. $P_-(m_z)$) the
probability for the grain to contain $N_0+1$ (resp. $N_0$) electrons,
it reads:
\begin{multline} 
\partial_t P_+=
%\frac{\partial P_+}{\partial t}=
   \sum_{a,\sigma}
  \frac{\Gamma_a^\sigma}{2} \Bigl(1+\eta\sigma m_z\Bigr) 
\biggl\{
n(\Delta E^{\sigma} -\mu_a) P_- 
% n(\Delta E^{\sigma}_a)) P_- 
\\ 
- \Bigl(1 - n(\Delta E^{\sigma} -\mu_a)\Bigr) P_+  
%- (1 - n(\Delta E^{\sigma}_a)) P_+  
\biggr\} 
\nonumber
\end{multline} 
where $\Delta E^{\sigma} (m_z)= \epsilon
-\kappa\sigma m_z -\hbar \gamma H \sigma/2$ is the energy to add an
electron in the nanomagnet with spin $\sigma$.  The offset
energy $\epsilon$ includes the charging energy, a 
possible capacitive coupling to an additional gate
and the energy of the level $\alpha_0$.  $\eta=1$ (resp. -1) when $\alpha_0$
is a majority (resp. minority) level.
On short time scales $\propto \hbar/\Gamma$, the charge equilibrate and $P_+/P_-$
reaches its stationary value. The latter is inserted into the first order part
of the $1/S_0$ expansion to obtain an equation for
$P(m_z)\equiv P_+(m_z)+P_-(m_z)$ of the form $\partial_{t} P(m_z) =
\partial_{m_z} [ -R(m_z) P(m_z)+\frac{1}{S_0} \partial_{m_z} \{D(m_z)  P(m_z)\} ]$.
This is a Fokker-Planck equation in the Ito form~\cite{GardinerBook}.
The associated Langevin equation for $m_z$ reads $\partial_t m_z= R(m_z) + \sqrt{2D(m_z)/S_0} \ \xi(t)$, where $\xi(t)$ is a white noise.
The noise part is small at large $S_0$ (the complete expression of $D(m_z)$ 
will be given in~\cite{ParcolletWaintalLong})
so that in this letter, we focus on the deterministic part $R(m_z)$. 
Defining the asymmetry parameter $q=(\sum_a \Gamma^\uparrow_a
-\Gamma^\downarrow_a)/(\sum_a\Gamma^\uparrow_a +\Gamma^\downarrow_a)$,
the total tunneling rate $\Gamma=\sum_a\Gamma^\uparrow_a
+\Gamma^\downarrow_a$ and the functions $f_{\sigma}(m_z)= \sum_a \Gamma_a^{\sigma} n
(\Delta E^{\sigma}-\mu_a)$, we obtain the equation governing the spin dynamics of the nanomagnet:
\begin{multline}
  \label{eq:rocks}
  \frac{\partial m_z}{\partial t} = (1-m_z^{2}) 
\biggl[
 \alpha_{B} \left( \gamma H + \frac{2\kappa}{\hbar} m_z\right) +
\\
\frac{(1-q) f_{\uparrow} (m_z) - (1+q)f_{\downarrow} (m_z)}{4S_{0}  ( 1 + q \eta m_z)}
\biggr] 
\end{multline}
where $\alpha_B\propto\Gamma_B$ is the Gilbert damping associated to the bosonic bath.
Equation (\ref{eq:rocks}) is the main result of this letter. 
%A similar equation can be derived for the current \cite{ParcolletWaintalLong}.

Let us now compare Eq. (\ref{eq:rocks}) with the LLG equation that has been  
widely used in multilayer systems,
 \be \label{eq:LLG} \frac{\partial\vec m}{\partial
t} = \vec m\times\left[ \frac{1}{\hbar S_0} \frac{\partial E}{\partial
\vec m} + \alpha \frac{\partial\vec m}{\partial t} + \frac{IP}{e S_0}
\frac{\vec z\times \vec m}{1+\bar q m_z} \right] \ee 
It consists of (i) a conservative term (the
magnetization precesses around trajectories of constant energy $E(\vec m)=- \kappa S_0 m_z^2 - \hbar\gamma S_0 H m_z$); 
(ii) a phenomenological Gilbert damping term $\alpha$ that allows the system to relax
to its equilibrium; (iii) a current induced spin torque
term~\cite{slonczewski1996},
written here assuming a magnetic left lead (with $\vec m_L \parallel z$) and 
a non-magnetic right lead. The torque is asymmetric and modulated by $0\le \bar q < 1$
($\bar q=0$ however in a tunneling junction~\cite{waintal2000,slonczewski2005}).
Using the cylindrical symmetry of Eq.(\ref{eq:LLG}), we obtain :  
%an equation for the $z$ component of $\vec m$ is extracted:
\be
\label{eq:llg-z} \frac{\partial m_z}{\partial t} = \frac{(1-m_z ^2)}{1+\alpha^2} 
\left[\alpha \left(\gamma H +\frac{2\kappa}{\hbar} m_z \right) +
\frac{P I/(e S_0)}{1+\bar q m_z } \right] \ee 
Even though Eq. (\ref{eq:rocks}) looks similar to Eq.(\ref{eq:llg-z}),
its third term is more complex leading to richer physics :
{ \it  i)} conducting channels for up/down electrons can open or close when $m_z$ varies : 
there is a {\sl magnetic blockade} induced by the anisotropy,
and the torque term varies abruptly with the bias voltage as the different channels open;
{ \it  ii)} this term is non-zero even at zero current; it is more than
the current-induced spin-torque effect, and contains in particular relaxations processes 
due to the leads \cite{tserkovnyak2002}.
We emphasize that while the
usual treatment of spin torque physics is done in two separate steps
(magnetization dynamics on one hand and transport properties on the
other hand), here we treat on the
same footing the transport and magnetic degrees of freedom.  Some
steps in this direction were taken previously by one of
us~\cite{waintal2003} for grains close to equilibrium.  

Let us now investigate the possibility of voltage-driven spin reversal. 
In the framework of the LLG equation, 
a current $I$ with a polarization $P$
flowing through a macrospin $S_0$ 
reverses the magnetization with a rate $P I/(e S_0)$ ($2S_0$ polarized
spins $+1/2$ are needed to fully reverse a large spin $S_0$ from
$S_z=-S_0$ to $S_z=+S_0$ provided the system ``absorbs'' all the
injected polarized spins). This torque term $P I/e S_0$ competes
with the relaxation rate $\alpha \kappa/\hbar$ (as opposed
to $\kappa$ alone in magnetic-field-driven reversals)
%here $\alpha$ is the Gilbert damping constant and $\kappa$ the uniaxial anisotropy energy per spin)
so that the reversal occurs for $\Theta_\text{LLG} \equiv P I \hbar/ (e S_0
\alpha\kappa)\sim 1$.  For example, for the nanopillars studied
in~\cite{katine2000}, $I= A j$ ($A$ cross section of the pillar and
$j$ current density) while $S_0=M_s A d /(\hbar\gamma)$ ($M_s$
saturation magnetization and $d$ thickness of the magnetic layer) 
leads to a critical value of the current
density $j_c= e M_s d \alpha\kappa/(P \hbar^2 \gamma)$ which gives the
observed value $j_c\sim 10^7 A.cm^{-2}$ for typical parameters.
In Eq. (\ref{eq:rocks}), the general condition for  voltage-driven spin reversal
(as a function $V_{\rm bias}$ and $H$) is $\Theta>2(1-q)$
where $\Theta \equiv \hbar \left[ (1-q)
\Gamma_L^\uparrow - (1+q) \Gamma_L^\downarrow \right]/(8\kappa
\alpha_B S_0)$ (corresponding to the value of the torque when all 
channels are open~\cite{ParcolletWaintalLong}).

\begin{figure}[htb]\includegraphics[width=8cm]{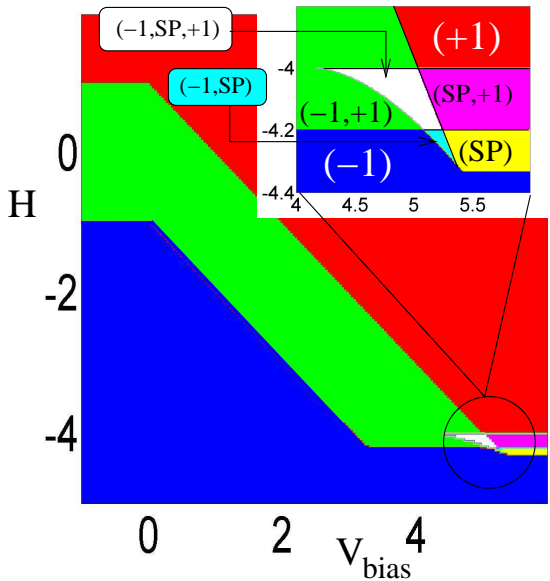}
\caption{ \label{fig:OneMagLead} (Color online). 
Phase diagram of Eq. (\ref{eq:rocks})  as a function of bias voltage $V_{\rm bias}$ 
and magnetic field $H$, measured respectively in unit of $\kappa/e$ and $2\kappa/\gamma$. The tunneling rates
are in unit of $8 S_0 \alpha_B \kappa/\hbar$ and  $V_{\rm bias}$ has been translated of the charging energy.
$k_B T=0.01\kappa$. The color code corresponds to the stable fixed points as indicated on the figure 
(e.g. $(-1,+1)$ means that  $m_z=\pm 1$ are stable). 
We consider one magnetic (left) and
one normal (right) lead: $\Gamma^\uparrow_{\rm left}=12$,
$\Gamma^\downarrow_{\rm left}=4$, $\Gamma^\uparrow_{\rm right}=\Gamma^\downarrow_{\rm right}=8$. SP denotes a stable Spin Precession state. 
 Inset: zoom of the main
figure.}
\end{figure}

\begin{figure}[htb]
\vglue +0.45cm
\includegraphics[width=8cm]{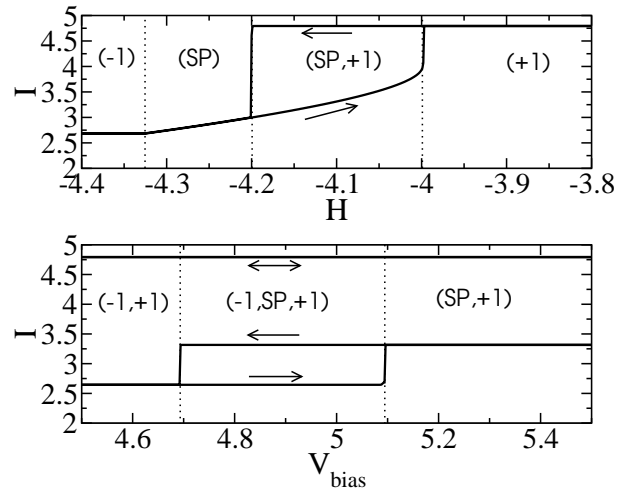}
\caption{\label{fig:current} Current as a function of bias voltage $V_{\rm bias}$ with $H=-4.1$ (lower panel) and
magnetic field $H$ with $V_{\rm bias}=5.5$ (upper panel) for the same system as Fig.~\ref{fig:OneMagLead}.}
\end{figure}

%{\it Phase diagram of the \nanoLLG equation.} 
The complete phase diagram of Eq. (\ref{eq:rocks}) as a function of $V_{bias}$ and $H$ 
%Eq.(\ref{eq:llg-z}) ~\cite{bazaliy2004} 
is found by studying its stable fixed points (i.e. the values of
$m_z$ satisfying $R(m_z)=0$ with $\partial_{m_z} R(m_z)<0$).
We restrict the following discussion to low temperature, $\eta=1$ and to the case
$f_R=0$ (Coulomb blockade situation, $\epsilon$ is the largest
energy). 
%When the two magnetic leads have their magnetizations
%aligned, the torque term vanishes altogether in Eq.(\ref{eq:rocks}) by
%symmetry, leading to a trivial phase diagram where the system
%switches when $H=\pm 2\kappa/\gamma$ independently of the value of
%$V_{\rm bias}$. 
Two identical magnetic leads with anti-parallel
magnetizations is the most favorable case for voltage-driven
switching: not only up electrons have a higher probability to enter
the grain than the down electrons, but in addition, once in, they have
a lower probability to get out so that the in-balance between up and
down current is enhanced. In this case however, the asymmetry parameter $q=0$
vanishes. The situation gets more interesting when $q\ne 0$ which is most easily
achieved by taking one lead to be magnetic while the other is not. The
torque can be strong enough to destabilize the $m_z=-1$ configuration
but since its magnitude decreases with increasing $m_z$, a stable
fixed point $-1<m^*_z<+1$ can appear. These fixed points are called
(uniform) spin precession states (SP) in the following.  A typical
phase diagram is given in Fig.~\ref{fig:OneMagLead} (where 
the different regions are labeled by the list of the stable
fixed points). In addition to
the regions that can exist in the LLG equation, $(SP)$,
$(SP,+1)$ and $(-1,SP)$, a region $(-1,SP,+1)$ where three states are
stable can also exist. In this region, the up spin channel is blocked
at $m_z=-1$ so that the torque term does not destabilize the $(-1)$
phase. For $m_z$ slightly higher, this channels opens up, and
stabilizes a spin precession fixed point.  The condition for observing
a $(SP,+1)$ and $(-1,SP,+1)$ region is $\Theta>(1-q)^2 /q $; the
condition for observing a $(SP)$ and $(-1,SP)$ region is
$\Theta>(1-q^2)/q$ \cite{ParcolletWaintalLong}.
This phase diagram is highly hysteretic as can be inferred from the measurement of the current $I$ flowing through the grain.  
In Fig.~\ref{fig:current} we show two hysteretic loops corresponding to
cross-sections of Fig.~\ref{fig:OneMagLead}. We note that in the lower panel, we find two different 
unconnected loops: if the system is initially in $m_z=+1$, it will stay there, no matter the value 
of $V_{\rm bias}$. On the contrary, if the system is initially in a SP state or in $m_z=-1$, 
it will switch, hysteretically between the $SP$ and the $-1$ state.
%
%
%%%%%%%%%%%%%%%%%%%%%%%%%%%%%%%%%%%%%%%%%%%%%%%%%%%%%%%%%%%%%%%%%%%%%%%%%%%%%%%%%%%%%%%%%
%{\it Discussion.}  

In actual nanometric cobalt grains, precise
measures of the anisotropy $\kappa$ have been obtained using a variety
of techniques (including microSQUID~\cite{jamet2001,jamet2004},
tunneling spectroscopy~\cite{gueron1999}, and
application of superparamagnetism~\cite{respaud1998}) and is of the order of
$\kappa/k_B\approx 100 mK$ (mainly due to surface anisotropy). Typical
currents~\cite{gueron1999} are of a few $nA$ which gives
$\Gamma\approx 10^{10} - 5.10^{10} s^{-1}$ while the total spin
$S_0\approx 1000$.  Typical polarizations are $P\sim 0.5-0.8$. More
difficult to estimate is the damping constant $\alpha_B$.  Switching
times were measured in Ref.\cite{thirion2003} using a microSQUID
technique combined with radio-frequency field pulses in the GHz range
leading to an estimated total damping constant of the order of $0.01$
for a $20 nm$ particle. A significant part of this damping is due to
spin relaxation through the lead and shall not be taken into account
in our estimation of the intrinsic damping $\alpha_B$ (in thin films
for instance, the adjunction of a metallic layer below and above a
magnetic layer is known to increase the damping by a factor $10$
due to spin relaxation through the conducting
electrons~\cite{back1999,tserkovnyak2002}).  Also, $20$ nm particles
probably have modes of relaxations that disappears for smaller
sizes. Hence, we estimate $\alpha_B\approx 0.001-0.005$ ($0.005$ being
the bulk value for Cobalt).  Putting everything together, we get
$\Theta \sim 1$.  This is of the order of the critical value for voltage-driven spin reversal
but is probably too small to excite spin precession states.  
Better candidates are magnetic alloys that would
allow to decrease the anisotropy (permalloy) and/or the total
magnetization (weak ferromagnets like NiPd should be good
candidates~\cite{kontos2002}) and increase the parameter $\Theta$ by
one or two orders of magnitude. Magnetic semiconductors (like GaMnAs)
could also be material of choice to study the effects described in
this letter.

\acknowledgments
We thank K. Mallick, F. Portier and P. Roche for useful comments.

\newcommand{{{\PRB}}}{{{Phys. Rev. B}}}\newcommand{{{\PRL}}}{{{Phys. Rev. Lett}}}\newcommand{{{\NPB}}}{{{Nucl. Phys.}}}\newcommand{{{\RMP}}}{{{Rev. Mod. Phys.}}}\newcommand{{{\ADV}}}{{{Adv. Phys.}}}

\end{document}